\documentclass[final,3p,times,twocolumn]{elsarticle}

\usepackage{amsthm}
\usepackage{amsopn}
\usepackage{amsmath}
\usepackage{url}
\usepackage{paralist}

\theoremstyle{plain}

\theoremstyle{definition}
\newtheorem{definition}{Definition}

\newcommand{\state}{s}
\newcommand{\states}{S}
\newcommand{\alphabet}{\Sigma}
\newcommand{\allprobs}{\mathcal{P}}
\newcommand{\initstate}{\state^{i}}

\newcommand{\reachset}{\mathcal{S}}
\newcommand{\errthres}{\epsilon}
\newcommand{\sys}{\mathcal{A}}
\newcommand{\prob}[3]{p_{{#1},{#2}}(#3)}
\newcommand{\computed}{v}
\newcommand{\smbol}{\alpha}
\newcommand{\smbolp}{\beta}
\newcommand{\plength}{n}
\newcommand{\statesi}{i}
\newcommand{\fpath}{\pi}
\newcommand{\ipath}{\rho}
\newcommand{\extension}[1]{{#1}^{\uparrow}}
\newcommand{\fword}{w}
\DeclareMathOperator{\lastWord}{last}
\newcommand{\last}[1]{\lastWord({#1})}
\newcommand{\infimum}{I}
\newcommand{\phit}{\mathcal{H}}
\newcommand{\pnhit}{\lnot\mathcal{H}}
\newcommand{\measset}{\mathcal{Z}}
\DeclareMathOperator{\lenWord}{len}
\newcommand{\len}[1]{\lenWord({#1})}
\newcommand{\iword}{\psi}
\DeclareMathOperator{\prWord}{Pr}
\newcommand{\lb}{l}
\newcommand{\ub}{u}
\newcommand{\iter}{r}
\newcommand{\recallEquation}[1]{Eq.~\ref{#1}}
\newcommand{\recallTable}[1]{Table~\ref{#1}}
\newcommand{\finmc}{\mathcal{M}}
\newcommand{\wprefix}[2]{\text{${#1}\!\!\downarrow\!\!{#2}$}}
\newcommand{\limword}{\vec{\iword}}
\newcommand{\recallInequation}[1]{Inequation~\ref{#1}}
\newcommand{\recallExternalThm}[1]{Thm.~{#1}}
\newcommand{\recallExternalDef}[1]{Def.~{#1}}
\newcommand{\st}{\mid}
\newcommand{\recallDefinition}[1]{Def.~\ref{#1}}
\newcommand{\bigiter}{R}
\newcommand{\iwordeps}{\iword^{\epsilon/2}}
\newcommand{\recallFigure}[1]{Fig.~\ref{#1}}

\newcommand{\qsep}{:}
\newcommand{\belief}[3]{b^{#1}_{#2}({#3})}
\newcommand{\beliefns}[2]{b^{#1}_{#2}}
\DeclareMathOperator{\supp}{supp}
\newcommand{\card}[1]{|{#1}|}
\newcommand{\sequ}{\sigma}
\DeclareMathOperator{\orderWord}{order}
\newcommand{\order}[1]{\orderWord(#1)}
\newcommand{\lsw}{LSW}
\newcommand{\lsws}{LSWs}
\DeclareMathOperator{\allwords}{Words}
\DeclareMathOperator{\lswWord}{LSW}
\newcommand{\alllsw}[1]{\lswWord({#1})}
\DeclareMathOperator{\statesWord}{States}
\newcommand{\stec}[1]{\statesWord({#1})}
\newcommand{\explanation}[1]{$\{$\mbox{#1}$\}$}
\newcommand{\pr}{\prWord\,\!}

\usepackage{amssymb}

\journal{Information Processing Letters}

\begin{document}

\begin{frontmatter}


\title{An algorithmic approximation of the infimum reachability
			probability for Probabilistic Finite Automata}
\author{Sergio Giro\corref{cor1}}
\address{FaMAF -- Universidad Nacional de C\'ordoba -- Argentina\\
			FCEIA -- Universidad Nacional de Rosario}

\cortext[cor1]{Urquiza 1949 16/F.
							Rosario, (2000) Rosario, Argentina}





\begin{abstract}
Given a Probabilistic Finite Automata (PFA), a set of states $\reachset$,
and an error threshold $\errthres > 0$, our algorithm approximates the
infimum probability (quantifying over all infinite words) that the automata
reaches $\reachset$. Our result contrasts with the known result that the
approximation problem is undecidable if we consider the supremum instead of the
infimum. Since we study the probability of reaching a set of states, instead of
the probability of ending in an accepting state, our work is more related to
model checking than to formal languages.
\end{abstract}


\end{frontmatter}


\section{Introduction}
\label{}

Suppose you want to analyse a system $\sys$ whose number of states is finite.
This system reacts to inputs from the environment in
a probabilistic fashion: if $\sys$ is in state $s$ and receives
$\alpha$ from the environment, the probability that $\sys$ transitions
to state $s'$ is $\prob{s}{\alpha}{s'}$. Moreover, assume that the environment
cannot observe the state of $\sys$ in order to choose the particular input
$\alpha$. The analysis you want to perform on this system is to calculate a
tight lower bound of the probability that the system achieves a certain goal,
no matter what the inputs are. For instance, inputs can model notifications
of the (un)availability of resources, and you might want to check that your
system sends a message with probability at least $0.8$, no matter what the
available resources are.

The problem in the paragraph above can be modelled using Probabilistic Finite
Automata (PFA)~\cite{DBLP:journals/iandc/Rabin63,DBLP:journals/ai/MadaniHC03}.
The assumption that inputs do not depend on the internal state of the
state of the input is central to assert that a PFA model adequately
reflects the behaviour of the system. In case the environment can observe
the state of $\sys$ to choose the particular input $\alpha$, the problem
can be modelled using Markov Decision Processes (MDP)~\cite{BOOK:Puterman}.

The usual semantics for PFA rely on the concept of \emph{acceptance},
by considering the set of finite words ending in an acceptance state
with probability greater than a given cut-point $\eta$. In contrast,
we focus on the concept of \emph{reachability}, and we are interested
on the probability with which each infinite word reaches some of the states
in a given set $\reachset$. In the realm of MDPs, both the supremum and
the infimum probability can be calculated in polynomial
time~\cite{DBLP:conf/fsttcs/BiancoA95}. In contrast, in the PFA setting
the supremum problem is undecidable~\cite{DBLP:journals/ai/MadaniHC03} for
both finite and infinite words~\footnote{Here, we consider only infinite words,
as the infimum probability over finite words is either $1$, if the initial
state of the system is in $\reachset$, or $0$, if it is not.}.
In fact, the supremum probability that $\sys$ reaches a state in
$\reachset$ cannot be even approximated algorithmically. This undecidability
result was the key to prove undecidability results for MDPs under partial
information~\cite{DBLP:conf/formats/GiroD07} as well as undecidability
for Probabilistic B\"uchi Automata~\cite{DBLP:conf/fossacs/BaierBG08}.

We present an algorithm to approximate the infimum probability that a PFA
$\sys$ reaches a set of states $\reachset$. Moreover, the computed value
$\computed$ is a lower bound of the infimum and, by performing a
sufficient number of iterations, we can ensure that it is as close to the
infimum as desired. Using the value $\computed$, we can answer our motivating
problem by stating that ``the probability that the goal is achieved is at
least $\computed$, no matter what the inputs are''. The fact that the value
$\computed$ is close to the infimum implies that the bound we provide is tight.

\section{Algorithm}

For our algorithm, we use the following definitions: a Probabilistic Finite
Automata (PFA) is a quintuple
$\sys = (\states,\alphabet,\allprobs,\initstate,\reachset)$, where
$\states$ is a finite set of states, $\alphabet$ is a set of symbols,
$\allprobs$ is a set of probability distributions on $\states$, comprising
one probability distribution $\prob{\state}{\alpha}{\cdot}$ for each
pair $(\state,\alpha)$ in $\states \times \alphabet$. The state $\initstate$
is called the \emph{initial state} of $\sys$, and $\reachset$ is a set of
\emph{hitting states}. We assume $\initstate \not\in \reachset$.

A finite path in $\sys$ is a sequence
\[ \fpath = \initstate.\smbol_{1}.\state_{1}.\cdots
								.\smbol_{\plength}.\state_{\plength} \]
where $\smbol_{\statesi} \in \alphabet$ and $\state_{\statesi} \in \states$
for all $\statesi$. Note that paths always start with the initial state
$\initstate$. We write $\len{\fpath}$ for $\plength$ and $\last{\fpath}$
for $\state_{\plength}$. In an analogous way to finite paths, infinite paths
are infinite sequences alternating symbols and states. The set of all infinite
paths having the finite path $\fpath$ as prefix is denoted by
$\extension{\fpath}$.

Given a word $\iword$ over $\alphabet$, let $\iword[k]$ denote the $k$-th
symbol in $\iword$. For every infinite word $\iword$ over $\alphabet$,
for every finite path $\fpath$, the probability
$\pr^{\iword}(\extension{\fpath})$ is defined as $1$
if $\fpath = \initstate$; if $\iword[\len{\fpath}+1] = \smbol$, we have
$\pr^{\iword}(\extension{\fpath.\smbol.\state})
	= \pr^{\iword}(\extension{\fpath})
				\cdot \prob{\last{\fpath}}{\smbol}{\state}$; if
$\iword[\len{\fpath}+1] \not= \smbol$, then
$\pr^{\iword}(\extension{\fpath.\smbol.\state}) = 0$.
In the same way as for Markov chains and MDPs (namely, by resorting to the
Carath\'eodory extension theorem), the previous definition for sets of the form
$\extension{\fpath}$ can be extended in such a way that, for all infinite
words $\iword$, the value $\pr^{\iword}(\measset)$ is defined for all
measurable sets $\measset$ of infinite paths.

Let $\phit$ be the set of all infinite paths $\ipath$ such that
some of the states in $\ipath$ is in $\reachset$. The amount we want to
approximate is $\infimum = \inf_{\iword} \pr^{\iword}(\phit)$. Note that
$\phit$ can be written as 
\begin{equation}
\label{eq:def-phit}
\phit = \biguplus_{\fpath \in C} \extension{\fpath} \; ,
\end{equation}
where $C$ is the set of all finite paths $\fpath$ such that $\last{\fpath}$
is the only state of $\fpath$ in $\reachset$.

In order to approximate $\infimum$, our algorithm iterates producing two
values in each iteration $r$. One of the values is a lower bound $\lb_{\iter}$
and the other one is an upper bound $\ub_{\iter}$. These bounds comply with:
\begin{eqnarray}
\lb_{\iter} & \leq & \lb_{\iter+1} \label{ineq:lb-increase} \\
\lb_{\iter} & \leq & \infimum \label{ineq:lb-islower} \\
\lim_{\iter \to \infty} \lb_{\iter} & = & \infimum \label{eq:lb-tends-inf} \\
\ub_{\iter} & \geq & \ub_{\iter+1} \label{ineq:ub-decrease} \\
\ub_{\iter} & \geq & \infimum \label{ineq:ub-isupper}  \\
\lim_{\iter \to \infty} \ub_{\iter} & = & \infimum \label{eq:ub-tends-inf} \; . 
\end{eqnarray}
To approximate $\infimum$ with error at most $\errthres$, the algorithm stops
when $\ub_{\iter} - \lb_{\iter} < \errthres$ (this is guaranteed to occur as
both $\ub_{\iter}$ and $\lb_{\iter}$ converge to the
same limit), and then returns $\lb_{\iter}$. Note that $\ub_{\iter}$ is also a
value with error less than $\errthres$ but, in order to give a safe lower bound
on the probability that a hitting state is reached, we use the pessimistic
value $\lb_{\iter} \leq \infimum$.

In the next subsections, we show how to calculate upper and lower bounds
complying with the desired properties.

\subsection{Lower bounds}

Let $\phit_{\iter} = \biguplus_{C_{\iter}} \extension{\fpath}$ where
$C_{\iter}$ is the set of paths such that $\last{\fpath}$ is the only state
of $\fpath$ in $\reachset$ and $\len{\fpath} \leq \iter$. By making the
same observation as for~\recallEquation{eq:def-phit}, we deduce that
$\phit_{\iter}$ is the set of all infinite paths reaching $\reachset$
after at most $\iter$ symbols. We often profit from the inclusion
\[ \phit_{\iter} \subseteq \phit_{\iter+1} \; . \]

We take $\lb_{\iter} = \inf_{\iword} \pr^{\iword}(\phit_{\iter})$.
Next, we show that this number can be calculated by brute force. 

Since only the first $\iter$ symbols are relevant, we need to consider each of
the finite words $\fword$ having exactly $\iter$ symbols. The truncation
operator $\wprefix{\iword}{\iter}$, that returns the prefix of $\iword$
having length $\iter$, will thus be quite useful in this subsection.
In addition, we use the notation $\pr^{\fword}(\phit_{\iter})$ to mean
$\pr^{\iword}(\phit_{\iter})$, where $\iword$ is any infinite word such
that $\wprefix{\iword}{\iter} = \fword$.

For each $\fword$ with $\len{\fword} = \iter$, we construct a finite Markov
chain $\finmc$. The procedure resembles the standard unfolding of a
probabilistic automaton (or an MDP) for a particular
adversary~\cite{thesis:segala}, and so we merely outline it. The states of
$\finmc$ are pairs $(\state,k)$ with $\state$ in $\states$
and $0 \leq k \leq \iter$. To describe $\finmc$ briefly, let's say that
the path
$\initstate.\smbol_{1}.\state_{1}.\cdots.\smbol_{\plength}.\state_{\plength}$
in $\sys$ maps to the path
$(\initstate,0).(\state_{1},1).(\state_{2},2).
						\cdots.(\state_{\plength},\plength)$ in $\finmc$.
For all $0 \leq k < \iter$, the probability of transitioning
from $(\state,k)$ to $(\state',k+1)$ is $\prob{\state}{\fword[k+1]}{\state'}$
(note that these probabilities depend on $\fword$). For simplicity, the states
$(\state,\iter)$ are stuttering. The initial state of $\finmc$ is
$(\initstate,0)$. The previous definitions for $\finmc$ imply that the
probabilities of the paths in $\sys$ having length at most $\iter$ coincide
with the probabilities of the corresponding paths in $\finmc$:
\[
\begin{array}{rcl}
& & \pr^{\fword}_{\sys}(\extension{\initstate.\smbol_{1}.\state_{1}.
		\cdots.\smbol_{\plength}.\state_{\plength}}) \\
	& = & \prob{\initstate}{\smbol_{1}}{\state_{1}} \cdot 
		\prod_{k=1}^{\plength-1} \prob{\state_{k}}{\smbol_{k+1}}{\state_{k+1}}\\
	& = & \pr_{\finmc}( \, (\initstate,0).(\state_{1},1).(\state_{2},2).
							\cdots.(\state_{\plength},\plength) \, ) \; .
\end{array}
\]
As a consequence, the probability that $w$ reaches $\reachset$ in
at most $\iter$ steps equals the probability that $\finmc$ reaches a state in
$\reachset \times \{ 0, \cdots, \iter \}$. The latter probability can be
calculated using standard techniques, as it poses a simple reachability
problem for finite Markov chains.

We have just showed that $\lb_{\iter}$ is computable. We still need to prove
that it complies with the properties we need so that our main algorithm works.
In order to prove \recallInequation{ineq:lb-increase}, we use the fact
that $\lb_{\iter} = \min_{\fword \in W_{\iter}} \pr^{\fword}(\phit_{\iter})$,
where $W_{\iter}$ is the set of all words of length $\iter$. Let $\fword^{*}$ be
$\arg \min_{\fword \in W_{\iter+1}} \pr^{\fword}(\phit_{\iter+1})$
and $\fword^{*-1}$ be $\wprefix{\fword^{*}}{\iter}$. The required inequality
$\lb_{\iter} \leq \pr^{\fword^{*}}(\phit_{\iter+1})$ follows since
$\lb_{\iter} = \min_{\fword \in W_{\iter}} \pr^{\fword}(\phit_{\iter})
		\leq \pr^{\fword^{*-1}}(\phit_{\iter})
		= \pr^{\fword^{*}}(\phit_{\iter})
		\leq \pr^{\fword^{*}}(\phit_{\iter+1})$,
where the last inequality holds since
$\phit_{\iter} \subseteq \phit_{\iter+1}$.

Next, we prove \recallInequation{ineq:lb-islower}. Let
$\mu = ( \, \iword(m) \, )_{m=1}^{\infty}$ be a sequence of
infinite words such that
$\lim_{m \to \infty} \pr^{\iword(m)}(\phit_{\iter}) = \infimum$
and the sequence $( \, \pr^{\iword(m)}(\phit_{\iter}) \, )_{m=1}^{\infty}$
is non-increasing (such a sequence exists by definition of infimum). Let
$\fword^{*}$ be a word of length $\iter$ that appears infinitely often in the
sequence $( \, \wprefix{\iword(m)}{\iter} \, )_{m=1}^{\infty}$ (this word
exists as the sequence is infinite, and there are finitely many words of
length $\iter$).

We prove \recallInequation{ineq:lb-islower} by proving
$\pr^{\fword^{*}}(\phit_{\iter}) \leq \infimum$. Suppose, towards a
contradiction, that $\pr^{\fword^{*}}(\phit_{\iter}) > \infimum$. Then,
by definition of $\mu$ there exists $\iword(p)$ in $\mu$ such that
$\pr^{\fword^{*}}(\phit_{\iter}) > \pr^{\iword(p)}(\phit_{\iter})
															\geq \infimum$.
Since $\fword^{*}$ appears infinitely often in
$( \, \wprefix{\iword(m)}{\iter} \, )_{m=1}^{\infty}$, there exists
$q > p$ such that $\wprefix{\iword(q)}{\iter} = \fword^{*}$. Since
the values $\iword(m)$ are non-increasing, we reach the following
contradiction:
$\pr^{\fword^{*}}(\phit_{\iter})
	> \pr^{\iword(p)}(\phit_{\iter})
	\geq \pr^{\iword(q)}(\phit_{\iter})
	= \pr^{\fword^{*}}(\phit_{\iter})$.

It remains to prove \recallEquation{eq:lb-tends-inf}. In
other to prove this equality, let
$\Psi$ be the sequence
\[ ( \, \Psi_{\iter} =  \arg \min_{\fword \in W_{\iter}}
											\pr^{\fword}(\phit_{\iter})
							\, )_{\iter=1}^{\infty} \]
(the set $W_{\iter}$ has been defined above). Note that
\begin{equation}
\label{eq:alt-def-lb}
\lb_{\iter} = \pr^{\Psi_{\iter}}(\phit_{\iter}) \; .
\end{equation} 
Given $\Psi$, we construct an infinite \emph{limit} word\footnote{We use the word
\emph{limit} as it resembles the \emph{limit schedulers}
in~\cite{DBLP:journals/entcs/GiroD09}.} $\limword$ having the property that,
for every $M$, the prefix $\wprefix{\limword}{M}$ appears infinitely often in
the sequence $( \, \wprefix{\Psi_{\iter}}{M} \, )_{\iter=M}^{\infty}$.
We take the first symbol $\limword[1]$ to be any symbol that appears
infinitely often in $( \, \wprefix{\Psi_{k}}{1} \, )_{k=1}^{\infty}$.
In order to obtain the second symbol $\limword[2]$, we consider
the subsequence $\Psi^{1}$ of $\Psi$ containing all words in $\Psi$ whose
first symbol is $\limword[1]$. Then, $\limword[2]$ is any symbol
that appears infinitely often as the second symbol in 
$( \, \wprefix{\Psi^{1}_{k}}{2} \, )_{k=2}^{\infty}$. In general, we can
describe the process to obtain $\Psi^{M}$ and $\limword[M]$
in an inductive fashion,  by stating that $\limword[M]$ is any symbol that
appears infinitely often in $( \, \Psi^{M-1}_{k}[M] \, )_{k=M}^{\infty}$ and
$\Psi^{M}$ is an (infinite) subsequence of $\Psi^{M-1}$ complying with
$\Psi^{M}_{k}[M] = \limword[M]$. The existence of the subsequence $\Psi^{M}$ 
ensures that $\wprefix{\limword}{M}$ appears infinitely often in
$( \, \wprefix{\Psi_{\iter}}{M} \, )_{\iter=M}^{\infty} \:$, as desired.

As an auxiliary result, we prove $\pr^{\limword}(\phit) = \infimum$.
Suppose, towards a contradiction, that $\pr^{\limword}(\phit) > \infimum$.
Then, there exists $\iword'$ such that
$\pr^{\limword}(\phit) > \pr^{\iword'}(\phit) \geq \infimum$.
As\footnote{This equality is standard for reachability properties, and can
be deduced from
$\pr^{\iword}(\phit)
= \pr^{\iword}(\biguplus_{k=1}^{\infty} \phit_{k} \setminus \phit_{k-1})
= \sum_{k=1}^{\infty} \pr^{\iword}(\phit_{k} \setminus \phit_{k-1})$.}
\begin{equation}
\label{eq:reach-by-limit}
\forall \iword \qsep \pr^{\iword}(\phit)
= \lim_{k \to \infty} \pr^{\iword}(\phit_{k}) \; ,
\end{equation}
there exists $K$ such that
\begin{equation}
\label{ineq:limword-gt-other-word}
\pr^{\limword}(\phit_{K}) > \pr^{\iword'}(\phit)
					\geq \pr^{\iword'}(\phit_{M}) =
						\pr^{\wprefix{\iword'}{M}}(\phit_{M})
\end{equation}
for all $M$.
By definition of $\limword$, there exists $M > K$ such that
$\wprefix{\Psi_{M}}{K} = \wprefix{\limword}{K}$. Then,
$\pr^{\limword}(\phit_{K})
= \pr^{\wprefix{\limword}{K}}(\phit_{K})
= \pr^{\wprefix{\Psi_{M}}{K}}(\phit_{K})
\leq \pr^{\Psi_{M}}(\phit_{M})
\leq \pr^{\wprefix{\iword'}{M}}(\phit_{M})$ (where the last inequality holds by
definition of $\Psi_{M}$) thus
contradicting~\recallInequation{ineq:limword-gt-other-word}.

Now we are ready to prove $\lim_{\iter \to \infty} \lb_{\iter} = \infimum$.
Since $\lb_{\iter} \leq \infimum$ for all $\iter$, we have
$\lim_{\iter \to \infty} \lb_{\iter} \leq \infimum$. Suppose, towards a
contradiction, that $\lim_{\iter \to \infty} \lb_{\iter} < \infimum$. Then, by
$\pr^{\limword}(\phit) = \infimum$ and~\recallEquation{eq:reach-by-limit},
there exists $K$ such that
\begin{equation}
\label{ineq:limit-lt-inf}
\lim_{\iter \to \infty} \lb_{\iter} < \pr^{\limword}(\phit_{K})
	= \pr^{\wprefix{\limword}{K}}(\phit_{K}) \; .
\end{equation}
By definition of $\limword$, there exists $M > K$ such that
$\wprefix{\Psi_{M}}{K} = \wprefix{\limword}{K}$.
Then, by~\recallEquation{eq:alt-def-lb}, we have
$\lim_{\iter \to \infty} \lb_{\iter}
\geq \pr^{\Psi_{M}}(\phit_{M})
\geq \pr^{\wprefix{\Psi_{M}}{K}}(\phit_{K})
= \pr^{\wprefix{\limword}{K}}(\phit_{K})$, which
contradicts~\recallInequation{ineq:limit-lt-inf}.

\subsection{Upper bounds}

For our upper bounds, we use \emph{lasso-shaped} words (\lsw). A \lsw\ is an
infinite word of the form
$\iword = \smbol_{1} \cdots \smbol_{K}
			( \smbolp_{1} \cdots \smbolp_{M} )^{\omega}$, in which the last $M$
in which the sequence of symbols $\smbolp_{1} \cdots \smbolp_{M}$ is looped
infinitely many times. The name \emph{lasso-shaped} is borrowed from the
counterexamples for LTL properties of B\"uchi automata, this name being used,
for instance, in~\cite{DBLP:conf/tacas/SchuppanB05}. Such counterexamples also
consist of a finite stem and a sequence that is looped infinitely many times.

In this paper, we restrict to \lsws\ with $M \leq 2^{\card{\states}}$ (recall
that $\states$ is the set of states of the PFA), and we say that $K$ is the
$\emph{order}$ of $\iword$, denoted by $\order{\iword}$. Note that, because of
our restriction on the length of the loop, the amount of \lsws\ with order at
most $K$ is finite.

We denote by $\alllsw{\iter}$ the set of all \lsw\ with order at most $\iter$.
The set of all infinite words is denoted by $\allwords$.

For upper bounds, we take
$\ub_{\iter} = \inf_{\iword \in \alllsw{\iter}} \pr^{\iword}(\phit)$.
Inequalities~\ref{ineq:ub-decrease} and~\ref{ineq:ub-isupper} follow from
$\alllsw{\iter} \subseteq \alllsw{\iter+1} \subseteq \allwords$.

The computability of $\ub_{\iter}$ follows in a similar way to that of
$\lb_{r}$: the amount of \lsws\ having order at most $\iter$ is finite, and we
can explore the probabilities for each of these words. Similarly as for the
lower bounds, the probability for a word $w_{1}(w_{2})^{\omega}$ is calculated
by constructing a finite Markov chain. We just outline the construction. The
set of the states of the Markov chain is
\newcommand{\isstem}{\mathsf{S}}
\newcommand{\isloop}{\mathsf{L}}
\[ \states
	\times \{ \; 1, \, \cdots, \, \max \{ K, M \} \; \}
	\times \{ \isstem, \isloop \} \]
(where $K = \len{w_{1}}$ and $M = \len{w_{2}}$).
The initial state is $(\initstate,1,\isstem)$. In the state $(s,n,\isstem)$
($(s,n,\isloop)$, resp.), the probability distribution for the next state is
determined by the $n$-th symbol in the stem (in the loop, resp.) In symbols, the
probability of transitioning from $(\state,n,\isstem)$ to
$(\state',n+1,\isstem)$ is $\prob{\state}{\fword_{1}[n]}{\state'}$ whenever
$n < K$. From $(\state,K,\isstem)$ to
$(\state',K,\isloop)$, the probability is
$\prob{\state}{\fword_{1}[K]}{\state'}$. The probabilities for the loop are
defined in a similar way: the only difference is that in a state
$(s,M,\isloop)$ in the end of the loop, we have that
$\prob{\state}{\fword_{2}[M]}{\state'}$ is the probability of transitioning to
$(s,1,\isloop)$ (that is, we return to the beginning of the loop). Note that
all the paths with positive probability are of the form
\[
\begin{array}{cl}
 & (\state_{1},1,\isstem) \cdots (\state_{K},K,\isstem) \\
\cdots & (\state_{K+1},1,\isloop) \cdots (\state_{K+M},M,\isloop) \\
\cdots & (\state_{K+iM+1},1,\isloop),\cdots,(\state_{K+iM+M},M,\isloop)
				\quad \cdots \; .
\end{array}
\]
Is is easy to see that the probability
$\pr^{w_{1}(w_{2})^{\omega}}(\phit)$ is the probability of reaching a state
$(s,n,l)$ such that $s \in \reachset$, and so the minimum probability for all
words of order at most $K$ can be obtained by constructing a Markov chain for
each of such words.

It remains to prove \recallEquation{eq:ub-tends-inf}.
If $\infimum = 1$, then $\ub_{\iter} = 1$ for all $\iter$, and so the equation
is trivial. From now on, we concentrate on the case $\infimum < 1$. In order to
prove that the limit is the infimum, it suffices to show that, for all
$\epsilon$, there exists $\bigiter$ such that
\begin{equation}
\label{ineq:desired-lim}
\inf_{\iword \in \alllsw{\bigiter}} \pr^{\iword}(\phit)
												< \infimum+\epsilon \; .
\end{equation}
We can indeed restrict to $\epsilon$ such that
\begin{equation}
\label{ineq:restrict-epsilon}
\epsilon < 1-\infimum \; .
\end{equation}
(Having proved the result for such values, the result also holds for the values
$\epsilon'$ such that $\epsilon' \geq 1-\infimum$, by taking $\epsilon$ such
that  $\epsilon = (1-\infimum)/2 < 1-\infimum \leq \epsilon'$ and hence
$\inf_{\iword \in \alllsw{\bigiter}} \pr^{\iword}(\phit)
					< \infimum+\epsilon < \infimum+\epsilon'$.)

We prove~\recallInequation{ineq:desired-lim} by showing that there exists
$\iword^{*} = \fword_{1}(\fword_{2})^{\omega}$ with
$\len{\fword_{2}} \leq 2^{\states}$ such that $\pr^{\iword^{*}}(\phit)
														< \infimum + \epsilon$,
By taking $\bigiter$ to be the order of $\iword^{*}$, we obtain
\recallInequation{ineq:desired-lim}, that is, the value $\ub_{\bigiter}$
is $\epsilon$-close to $\infimum$.

Let $\iwordeps$ be an infinite word such that
$\pr^{\iwordeps}(\phit) < \infimum + \epsilon/2$ (such a word exists by
definition of infimum). Using this word, we construct the word $\iword^{*}$
with the desired properties. For this construction, we focus on
the probability of \emph{not} reaching $\reachset$ (that is, the probability of
all infinite paths such that none of the states is in $\reachset$).
By definition of $\iwordeps$, we know that $\iwordeps$ does not reach
$\reachset$ with probability greater than $1 - \epsilon/2 - \infimum$; in
symbols:
\begin{equation}
\label{ineq:iwordeps-gt-epsilon}
\pr^{\iwordeps}(\pnhit) > 1 - \epsilon/2 - \infimum \; ,
\end{equation}
where $\pnhit$ is the complement of $\phit$, that is, the set of all infinite
paths $\ipath$ such that $\ipath[k] \not\in \reachset$ for all $k$.

Using $\iwordeps$, we define $\iword^{*}$ in such a way that
\begin{equation}
\label{ineq:iword-star-avoids-reach}
\pr^{\iword^{*}}(\pnhit) > 1 - \epsilon - \infimum
\end{equation}
and so $\pr^{\iword^{*}}(\phit) < \infimum + \epsilon$. The proof proceeds by
finding numbers $K$ and $M$ such that the first $K+M$ symbols of $\iword^{*}$
are the same as in $\iwordeps$. We name these symbols
$\smbol_{1}, \smbol_{2}, \cdots, \smbol_{K}, \smbolp_{1}, \smbolp_{2},
					\cdots, \smbolp_{M}$.
After these symbols, the word $\iword^{*}$ repeats
$\smbolp_{1}, \cdots, \smbolp_{M}$ indefinitely. This word is illustrated
in~\recallFigure{fig:counter-example-pfa}. The intuition behind the proof is
that there exists a set $Q_{1}$ of states such that, after exactly $K$ steps,
there is sufficiently high probability to be in $Q_{1}$, without hitting
$\reachset$ (in the figure, states in $\reachset$ are represented with crosses).
Moreover, if $Q_{i+1}$ ($Q_{1}$, respectively) is the set of all states that
can be reached after symbol $\smbolp_{i}$ ($\smbolp_{M}$, resp.) occurs in some
state in $Q_{i}$ ($Q_{M}$, resp.), then $Q_{i} \cap \reachset = \emptyset$ for
all $1 \leq i \leq M$.
\begin{figure}
\centering
\input{counter-example-pfa.pstex_t}
\caption{\label{fig:counter-example-pfa}Avoiding
										$\reachset$ with high probability}
\end{figure}
We find $K$, $M$ and show that $\iword^{*}$ complies with
\recallInequation{ineq:iword-star-avoids-reach}.

In order to obtain the required $K$, $M$, we profit from the fact that a PFA
according to our definition can be seen as a particular case of an MDP. For the
sake of completeness, we show how our definition for PFA matches the definition
of MDP in~\cite{thesis:deAlfaro}. If the MDP underlying a PFA $\sys$ is obvious
to the reader, then the rest of this paragraph can be safely skipped.
In~\cite{thesis:deAlfaro}, (\recallExternalDef{3.1}), an MDP $\Pi = (S,A,p)$
is defined by a set of states $S$, a set of actions $A(s)$ enabled at each
state $s$, and probabilities $p_{st}(a)$ of stepping from $s$ to $t$ using $a$,
for each $a \in A(s)$. When mapping a PFA $\sys$ to an MDP $\Pi$, the set of
states $S$ of $\Pi$ is the same set of states as in $\sys$; for each $s$ the
set $A(s)$ of actions enabled is the set $\alphabet$; the probabilities
$p_{st}(a)$ in~\cite{thesis:deAlfaro} are simply $\prob{s}{a}{t}$.

Using the MDP underlying $\sys$, we can resort to the \emph{end-component
theorem} (\cite[\recallExternalThm{3.2}]{thesis:deAlfaro}). In terms of PFA,
the definition of an end component is as follows.
\begin{definition}
\label{def:end-component}
An end component is a set $E \subseteq \states \times \alphabet$ such that for
every states $\state_{1} \not= \state_{\plength}$ in (a pair in) $E$ there
exists a path
$\state_{1}.\smbol_{1}.\state_{2}.\cdots.\smbol_{\plength-1}.\state_{\plength}$
such that $(\state_{k},\smbol_{k}) \in E$ and
$\prob{\state_{k}}{\smbol_{k}}{\state_{k+1}} > 0$ for all
$1 \leq k \leq \plength-1$. We write $\stec{E}$ for the set of states of $E$.
When no confusion arises, we simply write $\state \in E$ instead of
$\state \in \stec{E}$.
\end{definition}
Let $\mathcal{E}$ be the set of infinite paths
$\state_{1}.\smbol_{1}.\state_{2}.\cdots$ such that there exists $T$ such that
the set $\{ (\state_{t},\smbol_{t}) \st t > T \}$ is an end component. The
end-component theorem states that $\mathcal{E}$ has probability $1$ for all
words. The paths in $\mathcal{E}$ are said to \emph{end} in an end component.
Then, the set of paths that do not end in an end component (that is, the paths
for which no such $T$ exists) has probability $0$ for all words and, roughly
speaking, can thus be disregarded in probability calculations.

From now on, we are interested on the set $\mathcal{E}$ comprising all paths
ending in an end component. Now we show a partition for $\mathcal{E}$. For all
finite paths $\fpath$, end components $E$, let $Z(\fpath,E)$ be the set of all
infinite paths $\fpath.\smbol_{1}.\state_{2}.\smbol_{2}.\cdots$ such that
$(\smbol_{k},\state_{k+1}) \in E$ for all $k$. Next, we prove that the set
$\mathcal{E}$ is equal to
$\mathcal{E}' = \biguplus_{(\fpath,E) \in \mathcal{Z}} Z(\fpath,E)$
where $\mathcal{Z}$ is the set of all pairs $(\fpath,E)$ such that
$\fpath$ is either the trivial path $\initstate$, and $\initstate \in E$; or
$\fpath = \initstate.\cdots
			.\state_{\plength-1}.\smbol_{\plength}.\state_{\plength}$
and $(\state_{\plength-1},\smbol_{\plength}) \not\in E$ and $\state_{\plength}$
in $E$. In words, the last state/symbol pair is not in $E$, but the last state
is. Clearly, the inclusion $\mathcal{E}' \subseteq \mathcal{E}$ holds as the
paths in $Z(\fpath,E)$ end in $E$ for all $\fpath$, $E$. In order to prove the
inclusion $\mathcal{E} \subseteq \mathcal{E}'$, we prove that any path
$\iword \in \mathcal{E}$ is also in $\mathcal{E}'$. Since
$\iword \in \mathcal{E}$, there exists $T$ as
in~\recallDefinition{def:end-component}. Let's consider the minimum such $T$.
The existence of $T$ ensures that $\ipath$ has a prefix $\fpath$ after which
all the pairs state/symbol are in $E$. Moreover, since we are considering the
minimum $T$, either $\fpath$ is the trivial path $\initstate$, and $\initstate$
is in $E$; or the last state/symbol pair before $\last{\fpath}$ is not in $E$.
In summary, the fact that $R$ is minimum ensures that there exists
$(\fpath,E) \in \mathcal{Z}$ such that $\ipath \in Z(\fpath,E)$. It remains to
prove disjointness, that is, $Z(\fpath,E) \cap Z(\fpath',E') \not= \emptyset$
imply $(\fpath,E) = (\fpath',E')$. Suppose that there exists
$\ipath \in Z(\fpath,E) \cap Z(\fpath',E')$. The set of all state/symbol pairs
that appear infinitely often in $\ipath$ are all the pairs in $E$
(as $\ipath \in Z(\fpath,E)$), and the same goes for $E'$, thus yielding
$E = E'$. It remains to prove $\fpath = \fpath'$. We have that $\fpath$ and
$\fpath'$ are both a prefix of $\ipath$. Moreover, since we consider only
finite paths in which the last state/symbol pair is not in $E$, we have that
$\fpath$ is the smallest prefix such that after $\fpath$ all the state/symbol
pairs are in $E$, and the same holds for $\fpath'$. Then, both $\fpath$ and
$\fpath'$ have the same length, and so $\fpath = \fpath'$.

As a consequence of the partition we found, and the end-component theorem, for
all words $\iword$ we have
$ \pr^{\iword}(\Omega)
= \pr^{\iword}(\mathcal{E})
= \sum_{\fpath} \sum_{\{ E \st (\fpath,E) \in \mathcal{Z} \}}
										\: \pr^{\iword}(\, Z(\fpath,E) \,)$.
If a paths ends in an end component $E$ and does not hit $\reachset$, then no
prefix hits $\reachset$, and $E$ has no states in $\reachset$. Hence, for all
words $\iword$ we have
\[ \pr^{\iword}(\pnhit) =
	\sum_{\{ \fpath \st \fpath \cap \reachset = \emptyset \}}
		\sum_{\{ E \st (\fpath,E) \in \mathcal{Z}
								\land E \cap \reachset = \emptyset \}}
			 	\pr^{\iword}(\, Z(\fpath,E) \,) \; . \]
The outer sum ranges over all finite paths such that no state is in $\reachset$
(which we denote as $\fpath \cap \reachset = \emptyset$), and the inner sum
ranges over all end components $E$ such that the last state/action pair in
$\fpath$ is not in $E$, the last state is in $E$, and no state of $E$ is in
$\reachset$ (denoted by $E \cap \reachset = \emptyset$). In particular, for the
word $\iwordeps$ in~\recallInequation{ineq:iwordeps-gt-epsilon}, we have
$\pr^{\iwordeps}(\pnhit) = \sum_{\fpath \cap \reachset = \emptyset}
			\sum_{(\fpath,E) \in \mathcal{Z} \land
					E \cap \reachset = \emptyset}
			\pr^{\iwordeps}(\, Z(\fpath,E) \,) > 1 - \epsilon/2 - \infimum$.
Then, there exists a finite set
$\mathcal{B} \subseteq \{ \fpath \st \fpath \not\in \reachset \}$ such that
$\sum_{\fpath \cap \reachset = \emptyset}
			\sum_{(\fpath,E) \in \mathcal{Z}
						\land E \cap \reachset = \emptyset}
			 	\pr^{\iwordeps}(\, Z(\fpath,E) \,) >
									1 - \frac{3}{4}\epsilon - \infimum$.
Let $B = \max_{\fpath \in \mathcal{B}} \len{\fpath}$. For the sake of brevity,
let $\mathcal{V}$ be the set of all pairs $(\fpath,E)$ such that
$\fpath \cap \reachset = \emptyset$, and $\len{\fpath} \leq B$, and
$(\fpath,E) \in \mathcal{Z}$, and $E \cap \reachset = \emptyset$, and
$\pr^{\iwordeps}(\, Z(\fpath,E) \,) > 0$. Then,
\begin{equation}
\label{ineq:prob-bounded-non-reaching}
\sum_{(\fpath,E) \in \mathcal{V}}
	\pr^{\iwordeps}(\, Z(\fpath,E) \,) > 1 - \frac{3}{4}\epsilon - \infimum \; .
\end{equation}

Note that we can restrict to the pairs $(\fpath,E)$ such that
$\pr^{\iwordeps}(\, Z(\fpath,E) \,) > 0$, as the pairs with probability $0$
do not affect the sum.
In addition, by \recallInequation{ineq:restrict-epsilon}, we have
$1 - \epsilon - \infimum > 0$, and so in the sum in
\recallInequation{ineq:prob-bounded-non-reaching} there is at least one
positive summand $\pr^{\iwordeps}(\, Z(\fpath,E) \,)$.

\begin{figure}
\centering
\input{exiting-end-component.pstex_t}
\caption{\label{fig:exiting-end-component}$(\state_{1},\smbol)$ is
					in $E$, but $(\state_{2},\smbol)$ is not}

\end{figure}
The desired $K$, $M$ are now obtained from $\iwordeps$ and $B$. Note that,
although we restricted to the summands complying with
$\pr^{\iwordeps}(\, Z(\fpath,E) \,) > 0$, it is still possible that $\sys$
exits $E$ after $\fpath$ with positive probability (as the same symbol might
be inside $E$ for a reachable state $\state$, but outside $E$ for a state
$\state'$ that is reachable after the same number of steps as $\state$, see
\recallFigure{fig:exiting-end-component}). We
show that, by considering arbitrarily large paths, the probability that $E$ is
exited becomes arbitrarily small.

Let $\belief{\fword}{k}{\state}$ be the probability that, after $k$ steps,
the state is reached is $\state$. We generalize this notation to sets of states.
Formally:
$\belief{\fword}{k}{T} = \sum_{\len{\fpath} = k, \last{\fpath} \in T}
											\pr^{\fword}(\extension{\fpath})$.
We call the distribution $\belief{\fword}{k}{\cdot}$ a \emph{belief state},
following the nomenclature for POMDPs~\cite{DBLP:journals/ai/KaelblingLC98}.
Since the set of states is finite, there exist two indices $x < y$ such that
$\supp(\beliefns{\iwordeps}{x}) = \supp(\beliefns{\iwordeps}{y})$
(where $\supp$ denotes the support of the distribution). Moreover, given any
two numbers $X$, $Y$, such that $Y > X+2^{\card{\states}}$, we have
$X \leq x \leq y \leq Y$ and
$\supp(\beliefns{\iwordeps}{x}) = \supp(\beliefns{\iwordeps}{y})$ for some
$x$, $y$. Since the amount of sequences of the form
$T_{0} \gamma_{1} \cdots \gamma_{V} T_{V}$ with $V \leq 2^{\card{\states}}$ is
finite (where each $T_{v}$ is a set of states), at least one of such finite
sequences appears infinitely many times in the infinite sequence
$\supp(\beliefns{\iwordeps}{0}) \iwordeps(1)
			\supp(\beliefns{\iwordeps}{1}) \iwordeps(2) \cdots$. Suppose this
finite sequence is $\sequ = T_{0} \gamma_{1} \cdots \gamma_{V} T_{V}$.

We show that we can take
$\beta_{1}, \cdots, \beta_{M}
		= \gamma_{1}, \cdots, \gamma_{V}$.
In addition, we take $K$ to be a number (defined below) greater than $B$, in
which an occurrence of $\sequ$ starts.

Given a component $E$ in a pair in $\mathcal{V}$ (defined before
\recallInequation{ineq:prob-bounded-non-reaching}), let
\begin{multline*}
Q^{E} =
	\{ \state' \in T_{0} \cap \stec{E} \st \forall v \leq V \qsep \\
				\pr^{\gamma_{1} \cdots \gamma_{v}}
				(\state'.\gamma_{1}.\state_{1}.\cdots.\gamma_{v}.\state_{v})
																	> 0
					\implies \state_{k} \in E \} \; .
\end{multline*}
In other words, $Q^{E}$ comprises the states in $T_{0} \cap \stec{E}$ from
which, when executing $\gamma_{1} \cdots \gamma_{V}$, we can only reach states
in $E$. Let $Q^{\lnot E}$ be $\stec{E} \setminus Q^{E}$.

Consider the infinite sequence
$e(1), e(2) \cdots$ of indices where $\sequ$ starts. We show that
$\lim_{v \to \infty} \belief{\iwordeps}{e(v)}{\state} = 0$ for all
$\state \in Q^{\lnot E} > 0$. (As the number of states is finite, this implies
$\lim_{v \to \infty} \belief{\iwordeps}{e(v)}{Q^{\lnot E}} = 0$.)
Suppose, towards a contradiction, that for some $s \in Q^{\lnot E}$, $l > 0$,
we have $\belief{\iwordeps}{e(v)}{s} \geq l$ for all $v$. By definition of
$Q^{\lnot E}$, there exists $\state' \not\in E$, $d > 0$, and $v$ such that
$\pr^{\gamma_{1} \cdots \gamma_{v}}
				(\state.\gamma_{1} \cdots \gamma_{v}.\state')
													= d$. Then, the
probability of staying in $E$ after the $n$-th repetition of $\sequ$ is
less than or equal to $(1-(l \cdot d))^{n}$, for all $n$. This implies that the
probability of staying in $E$ indefinitely is $0$, thus contradicting the fact
that $\pr^{\iwordeps}(\, Z(\fpath,E) \,) > 0$.

As a result, for all pairs $(\fpath,E)$ in $\mathcal{V}$, there exists
$e(\fpath,E) \in \{ e(1), e(2), \cdots \}$ such that
\begin{equation}
\label{ineq:escape-small}
\belief{\iwordeps}{e(\fpath,E)}{Q^{\lnot E}}
							< \epsilon/(4 \cdot \card{\mathcal{V}}) \; .
\end{equation}
Define
$K = \max(B , \{ e(\fpath,E) \st (\fpath,E) \in \mathcal{V} \})$
and 
$Y(\fpath,E) = Z(\fpath,E) \setminus
			\{ \ipath \st \ipath[K] \in Q^{\lnot E} \}$.
We have
$Z(\fpath,E) \subseteq Y(\fpath,E) \cup
				\{ \extension{\fpath} \st \fpath[K] \in Q^{\lnot E}
							\land \len{\fpath} = K \}$.
Then,
\begin{equation}
\label{ineq:bound-Z}
\pr^{\iword}(\, Z(\fpath,E) \,) \leq \pr^{\iword}(\, Y(\fpath,E) \,)
								+ \belief{\iword}{K}{Q^{\lnot E}}
\end{equation}
for all infinite words $\iword$.

We have
\begin{equation}
\label{ineq:star-better-than-eps-for-Y}
 \pr^{\iword^{*}}(\, Y(\fpath,E) \,) \geq \pr^{\iwordeps}( \, Y(\fpath,E) \,)
\end{equation}
as, under $\iword^{*}$, all paths of length $K$ ending in a state in
$Q^{E}$ do not reach states outside $E$ (because of our definition of $Q^{E}$
and the symbols $\gamma_{v}$). In fact, if $\len{\fpath} \geq K$, the
scenario in \recallFigure{fig:exiting-end-component} is possible under
$\iwordeps$, but not possible under $\iword^{*}$. Roughly speaking, after $K$
steps the word $\iword^{*}$ does not escape $E$, thus yielding higher
(or equal) probability for $Y(\fpath, E)$ than any word $\iword$ such that
$\wprefix{\iword}{K} = \wprefix{\iword^{*}}{K}$ and, in particular, than
$\iwordeps$. Then,
\[
\begin{array}{cl}
& \pr^{\iword^{*}}(\pnhit) \\
\geq & \explanation{$Y(\fpath,E)\!\subseteq\!Z(\fpath,E)$,
						the sets $Z(\fpath,E)$ partition $\pnhit$} \\
     & \sum_{(\fpath,E) \in \mathcal{V}} \pr^{\iword^{*}}(\, Y(\fpath,E) \,) \\
\geq & \explanation{Ineq.~\ref{ineq:star-better-than-eps-for-Y}} \;
     	\sum_{(\fpath,E) \in \mathcal{V}} \pr^{\iwordeps}(\, Y(\fpath,E) \,) \\
\geq & \explanation{Ineq.~\ref{ineq:bound-Z}} \;
		\sum_{(\fpath,E) \in \mathcal{V}}
			\pr^{\iwordeps}(\, Z(\fpath,E) \,)
			- \belief{\iwordeps}{K}{Q^{\lnot E}} \\
> & \explanation{Inequations~\ref{ineq:prob-bounded-non-reaching},
								\ref{ineq:escape-small}} \;
  	1 - \frac{3}{4}\epsilon - \infimum - \card{\mathcal{V}} \cdot
								\epsilon/(4\!\cdot\!\card{\mathcal{V}}) \\
\geq & 1 - \epsilon - \infimum
\end{array}
\]
In conclusion, the word
$\iword^{*}
	= \smbol_{1} \cdots \smbol_{K} (\smbolp_{1} \cdots \smbolp_{M})^{\omega}$
(where $\smbol_{1} \cdots \smbol_{K} \smbolp_{1} \cdots \smbolp_{M}$ are first
$K+M$ symbols in $\iwordeps$) complies with
\recallInequation{ineq:iword-star-avoids-reach}. Since
$\order{\iword^{*}} = K$, we obtain
$\inf_{\iword \in \alllsw{K}} \pr^{\iword}(\phit) < \infimum + \epsilon$. By
\recallInequation{ineq:ub-decrease}, this inequality implies
$\inf_{\iword \in \alllsw{k}} \pr^{\iword}(\phit) < \infimum + \epsilon$
for all $k \geq K$, thus ensuring \recallEquation{eq:ub-tends-inf}.

\section{Discussion}

Our algorithm is nonprimitive recursive, and we have still nothing to say
about the complexity of the problem.

However, the fact that there exists an algorithm to approximate the value is
quite surprising considering similar problems for PFA, as shown in
\recallTable{tab:decidability}. The table indicates, for the problems of
reachability and acceptance, whether there exists an
algorithm to approximate and/or to compute extremal values.
\begin{table}
\centering
\footnotesize
\begin{tabular}{|c|c|c|}
\hline
Approximate/Compute &  Infimum & Supremum \cite{DBLP:journals/ai/MadaniHC03} \\
\hline
Reachability 		&	$\surd / ?$	& $\times /	\times$
									\\
\hline
Acceptance	\cite{DBLP:journals/mst/BlondelC03} & $? / \times$
												& $\times / \times$   \\
\hline
\end{tabular}
\caption{\label{tab:decidability}Existence of algorithms for PFA}
\end{table}
Note that the only $\surd$ in the table corresponds to the result in this
paper. The table also indicates two pending questions: whether there exists
an algorithm to effectively compute the infimum for reachability, and whether
the infimum for acceptance can be approximated.

The undecidability for the supremum probability has been used to prove that
quantitative model checking under partial
information~\cite{DBLP:conf/formats/GiroD07,DBLP:conf/sbmf/Giro09} is
undecidable for properties involving the supremum. The setting of these papers
is more general, as several entities might have different information about the
state of the system (in contrast, the problem we address in this paper concerns
only an environment that has no information about the state of the system).
However, we expect that the proof we presented sheds some light on whether this
more general problem is computable or not, in case we consider the infimum
instead of the supremum.

\bibliographystyle{elsarticle-num}
\bibliography{probability.bib}

\end{document}